\documentclass[twocolumn,amsmath,amssymb]{revtex4-2}
\usepackage{dcolumn} 
\usepackage{graphicx} 
\usepackage{bm} 
\usepackage{braket} 
\usepackage[
colorlinks=true, 
citecolor=blue, 
urlcolor=blue, 
linkcolor=blue,
setpagesize=false, 
bookmarks=false
]{hyperref}

\begin{document}


\title{Ground-State Properties of the $t$-$J$ Model for the CuO Double-Chain Structure}
\author{Tatsuya Kaneko,$^{1}$ Satoshi Ejima,$^2$ Koudai Sugimoto,$^{3}$ and Kazuhiko Kuroki$^{1}$}
\affiliation{
$^1$Department of Physics, Osaka University, Toyonaka, Osaka 560-0043, Japan \\
$^2$Institut f\"ur Softwaretechnologie, Abteilung High-Performance Computing, Deutsches Zentrum f\"ur Luft- und Raumfahrt (DLR), 22529 Hamburg, Germany \\
$^3$Department of Physics, Keio University, Yokohama, Kanagawa 223-8522, Japan
}
\date{\today}


\begin{abstract}
We investigate the ground-state properties of a correlated model for the double-chain structure in cuprates. 
We consider the $t$-$J$ model, in which the nearest-neighbor spin interaction $J_1$ is smaller than the next-nearest-neighbor interaction $J_2$ corresponding to the CuO double-chain structure.   
We vary $J_1$ from antiferromagnetic to ferromagnetic values and calculate the correlation functions including the superconducting pair correlation function. 
Employing the density-matrix renormalization group method, we show that the ground state for antiferromagnetic $J_1$ exhibits the hallmarks of the Luther--Emery liquid phase, in which the spin-singlet pair and charge-density-wave correlations exhibit power-law decays against distance, and the spin correlation function decays exponentially. 
Its signatures are gradually dismissed as $J_1$ approaches the ferromagnetic regime.  
Our findings suggest that the antiferromagnetic double-chain structure without ferromagnetic bonds is favorable for superconductivity. 
\end{abstract}

\maketitle


\section{Introduction}

Low-dimensional correlated electron systems are one of the most attractive research fields in condensed matter physics. 
One representative example is high-temperature superconductivity (SC) in cuprates. 
The unconventional superconducting properties of cuprates have been investigated in the doped Mott insulators described by the Hubbard and $t$-$J$ models~\cite{dagotto1994,imada1998,lee2006,keimer2015}. 
Density-matrix renormalization group (DMRG) calculations enable us to obtain nearly unbiased numerical results, which have recently revealed, e.g., robust $d$-wave-like SC on a square lattice~\cite{jiang2019,gong2021} and possible topological SC on a frustrated triangular lattice~\cite{jiang2020,huang2022,huang2023}. 
In systems belonging to one-dimensional (1D) structures (e.g., cylinders used in these DMRG simulations), a hallmark of SC is given by the emergence of the Luther--Emery (LE) liquid state~\cite{luther1974}, which has the gapped-spin and gapless-charge modes with quasi-long-range SC pair correlations.  
Although the SC correlation competes with the charge-density-wave (CDW) correlation, a finding of the LE liquid phase may provide helpful insights into the material design of unconventional superconductors. 

SC on the CuO$_2$ layer, which consists of corner-sharing square-planar units, has been investigated extensively.  
On the other hand, cuprates {containing edge-sharing square-planar units also have the potential to show SC. 
For example, Pr$_2$Ba$_4$Cu$_7$O$_{15-\delta}$ composed of multiple copper oxide chain structures shows SC with a moderate number of oxygen defects~\cite{matsukawa2004,yamada2005,hagiwara2006,ishikawa2007,wakisaka2008,chiba2013,tada2013,kuwabara2016,taniguchi2021,hagawa2024}, where a double-chain structure (see Fig.~\ref{fig1}) is expected to be responsible for SC~\cite{sano2005,nakano2007,sano2007,okunishi2007,berg2007,nishimoto2008,habaguchi2011,nishioka2022}.  
The electronic properties in the edge-sharing structure have crucial differences from those in the corner-sharing structure. 
In the double-chain structure shown in Fig.~\ref{fig1}, whereas the hopping $t_2$ between next-nearest-neighboring (NNN) Cu sites is composed of the Cu-O-Cu ($d$-$p$-$d$) process as in the corner-sharing structure, the hopping $t_1$ between nearest-neighboring (NN) Cu sites has no $d$-$p$-$d$ contribution when the $p_x$ and $p_y$ orbitals in the shared O site are orthogonal. 
Hence, $|t_2| > |t_1|$~\cite{nakano2007,sano2007}, i.e., two CuO chains are weakly connected via a small $t_1$. 

\begin{figure}[b]
\begin{center}
\includegraphics[width=0.75\columnwidth]{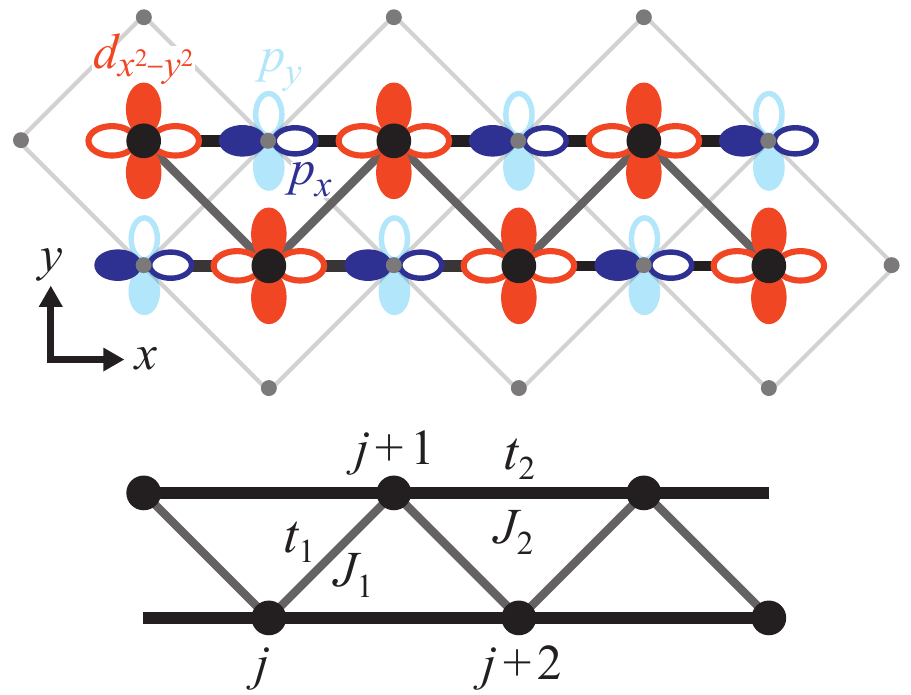}
\caption{CuO double-chain structure and zigzag chain.}
\label{fig1}
\end{center}
\end{figure}

As for magnetism, the double-chain (zigzag chain) model at half-filling can be mapped to the frustrated $J_1$-$J_2$ Heisenberg chain~\cite{nishimoto2008}. 
Because edge-sharing NN clusters have the ferromagnetic contribution $J^{(\rm FM)}_1$ mediated by the spin interaction at the ligand site~\cite{rice1993,arita1998,streltsov2017,autieri2022}, the Heisenberg model with ferromagnetic $J_1$ and antiferromagnetic $J_2$ for quasi-1D cuprates has been studied in detail~\cite{drechsler2007,hikihara2008,sudan2009,furukawa2012,agrapidis2019,yamaguchi2020}. 
In the double-chain structure shown in Fig.~\ref{fig1}, the superexchange mechanism attributed to the straight-line $d$-$p$-$d$ process gives a large antiferromagnetic $J_2$.    
On the other hand, the direct $d$-$d$ and detouring $d$-$p$-$p$-$d$ processes (included in $t_1$) can lead to the antiferromagnetic contribution $J^{(\rm AFM)}_1$. 
Hence, the competing $J^{(\rm AFM)}_1$ and $J^{(\rm FM)}_1$ may result in a small $J_1$ $(=J^{(\rm AFM)}_1 + J^{(\rm FM)}_1)$, and $J_2 > J_1$ is expected in the zigzag chain structure.  
The magnetism and possible SC in the CuO double-chain structure have been investigated in the $t_1$-$t_2$ Hubbard model~\cite{nakano2007,nishimoto2008}. 
However, the simple $d$-orbital-based Hubbard model does not consider the ferromagnetic contribution $J^{(\rm FM)}_1$ attributed to the ligand $p$ orbitals. 
Although possible SC has also been studied in the $d$-$p$ model and its effective $t$-$J$ model~\cite{sano2005,sano2007,berg2007}, long-range properties of correlation functions have not been explicitly investigated in large systems including the case of ferromagnetic $J_1$. 

In this paper, we investigate the ground-state properties of the double-chain structure employing the DMRG method. 
In particular, we consider the $t$-$J$ model to incorporate the ferromagnetic contribution in $J_1$.
Our calculations of the correlation functions including the SC pair correlation function show that the ground state for antiferromagnetic $J_1$ exhibits the hallmarks of the LE liquid phase, whereas the signatures of the LE liquid gradually vanish as $J_1$ approaches the ferromagnetic value. 
Our DMRG calculations provide insights into the SC state possibly emerging on the CuO double-chain structure. 

The rest of this paper is organized as follows.  
In Sect.~\ref{sec:model}, we introduce the model and method. 
In Sect.~\ref{sec:results}, we present the results of our DMRG calculations, where we show the spin, charge, and pair correlation functions for various $J_1$ values.  
Our conclusions are presented in Sect.~\ref{sec:summary}.


\section{Model and Method} \label{sec:model}

The Hamiltonian of the $t$-$J$ model on the zigzag chain structure shown in Fig.~\ref{fig1} reads 
\begin{align}
\hat{\mathcal{H}} =
& -t_1 \sum_{j,\sigma} \left( \hat{\tilde{c}}^{\dag}_{j,\sigma} \hat{\tilde{c}}_{j+1,\sigma} \!+\! {\rm H.c.}\right)
 -t_2 \sum_{j,\sigma} \left( \hat{\tilde{c}}^{\dag}_{j,\sigma} \hat{\tilde{c}}_{j+2,\sigma} \!+\! {\rm H.c.}\right)
 \notag \\
 &+J_1 \sum_{j} \left( \hat{\bm{S}}_{j} \cdot \hat{\bm{S}}_{j+1}  - \frac{1}{4}  \hat{n}_j \hat{n}_{j+1} \right)
 \notag \\
& +J_2 \sum_{j} \left( \hat{\bm{S}}_{j} \cdot \hat{\bm{S}}_{j+2}  - \frac{1}{4}  \hat{n}_j \hat{n}_{j+2} \right). 
\end{align}
$\hat{\tilde{c}}_{j,\sigma}=\hat{c}_{j,\sigma}(1-\hat{n}_{j,\bar{\sigma}})$ is the constrained annihilation operator of $\hat{c}_{j,\sigma}$ for a fermion with spin $\sigma$ ($=\uparrow,\downarrow$) at site $j$, where $\hat{n}_{j,\sigma}=\hat{c}^{\dag}_{j,\sigma}\hat{c}_{j,\sigma}$ and $\bar{\sigma}$ denotes the opposite spin of $\sigma$. 
The constrained hopping using the operators $\hat{\tilde{c}}^{\dag}_{j,\sigma}$ and $\hat{\tilde{c}}_{j,\sigma}$ prohibits the creation of a doubly occupied site. 
$\hat{\bm{S}}_j$ is the spin operator, and $\hat{n}_{j}$ ($= \sum_{\sigma} \hat{n}_{j,\sigma}$) is the number operator. 
$t_1$ and $t_2$ are the transfer integrals of the NN and NNN hoppings, respectively. 
$t_1$ ($t_2$) corresponds to the interchain (intrachain) hopping (see Fig.~\ref{fig1}). 
$J_1$ and $J_2$ are the NN and NNN spin interactions, respectively. 
In the $L$ site zigzag chain, the filling of the $N$-particle system is defined as $n=N/L=(N_{\uparrow}+N_{\downarrow})/L$, where $N_{\sigma}$ is the number of spin $\sigma$. 

We consider the $t$-$J$ model underlying the $d$-$p$ structure in the CuO double chains, e.g., in Pr$_2$Ba$_2$Cu$_7$O$_{15-\delta}$~\cite{sano2005,sano2007,nakano2007}.   
Although the electron filling of the double-chain structure in Pr$_2$Ba$_4$Cu$_7$O$_{15-\delta}$ has not been specified in experiments, it is expected to be slightly above quarter-filling ($n=0.5$)~\cite{sano2005,nakano2007}. 
Hence, in this paper, we set $n = 0.6$. 
The hopping $t_2$ between NNN Cu sites is larger than $t_1$ between NN Cu sites because the $d_{x^2-y^2}$-$p_x$--$d_{x^2-y^2}$ network leads to a large effective $d$-$d$ hopping for $t_2$ but it is absent in $t_1$ (see Fig.~\ref{fig1}). 
The superexchange in the strong $d_{x^2-y^2}$-$p_x$-$d_{x^2-y^2}$ bond leads to a large antiferromagnetic NNN interaction $J_2$ ($>0$).
On the other hand, the direct $d$-$d$ hopping and detouring $d$-$p$-$p$-$d$ processes can yield the antiferromagnetic NN contribution $J_1^{(\rm AFM)}$~($>0$).  
The NN interaction $J_1$ also includes the ferromagnetic contribution $J_1^{(\rm FM)}$~$(<0)$ because of Hund's coupling between the $p_x$ and $p_y$ orbitals at a shared O site~\cite{rice1993,arita1998,streltsov2017,autieri2022}. 
Hence, $J_1 =  J_1^{(\rm AFM)} + J_1^{(\rm FM)}$, i.e., the antiferromagnetic and ferromagnetic contributions are competing in the NN interaction. 
If $|J_1^{(\rm FM)}| > J_1^{(\rm AFM)}$, $J_1$ can be negative.   
In this paper, assuming $|t_2| > |t_1|$ and $J_2 > J_1$, we investigate the ground-state properties at various $J_1$ values. 
We set $J_2$~($>0$) as the unit of energy and use $t_2 = 3J_2$. 
Referring to the values of hoppings used in previous studies~\cite{nakano2007,sano2007}, we set $t_1/t_2$ = 0.2 ($t_1 =0.6J_2$). 

We employ the DMRG method~\cite{white1992,white1993} to obtain the ground state in the $t$-$J$ model. 
We apply open boundary conditions (OBCs) to the $L$ site zigzag chain and perform DMRG calculations in the $S^z_{\rm tot}=0$ (i.e., $N_{\uparrow}=N_{\downarrow}$) sector. 
The bond dimension is up to $m= 10000$, where the largest truncation error is on the order of 10$^{-7}$.  
As discussed in Appendix~\ref{appendix_A}, the ground state at $J_1 < 0$ can be a $\braket{\hat{\bm{S}}^2_{\rm tot}}=S_{\rm tot}(S_{\rm tot}+1) > 0$ state (where $\hat{\bm{S}}_{\rm tot}=\sum_j \hat{\bm{S}}_j$) depending on the system size and boundary conditions.  
In this paper, we address the parameter regime, in which the ferromagnetic contribution to $J_1$ is small and small differences in $S_{\rm tot}$ do not change our main conclusion. 
Unless otherwise noted, we show the results using the $L=200$ site zigzag chain at $n=0.6$, in which $S_{\rm tot}=0$ under OBCs. 

To assess ground-state properties, we compute the local charge density $n(j) = \braket{\hat{n}_j}$ and three types of correlation function. 
The charge-density and spin correlation functions are defined as 
\begin{align}
C(r) = \braket{\hat{n}_{j_0} \hat{n}_{j_0+r}} - \braket{\hat{n}_{j_0}}  \braket{\hat{n}_{j_0+r}}, 
\end{align}
and
\begin{align}
S(r) = \braket{\hat{S}^{z}_{j_0} \hat{S}^{z}_{j_0+r} }, 
\end{align}
respectively, where $j_0$ is the reference site. 
In addition, to obtain pairing properties, we calculate the pair correlation function 
\begin{align}
P (r) = \braket{ \hat{\Delta}^{\dag}_{j_0} \hat{\Delta}_{j_0+r}}, 
\end{align}
where $\hat{\Delta}_j$ is the spin-singlet pair operator defined as $\hat{\Delta}_j =  \left( \hat{c}_{j,\uparrow} \hat{c}_{j+1,\downarrow} - \hat{c}_{j,\downarrow} \hat{c}_{j+1,\uparrow}\right) / \sqrt{2}$. 
In the model shown in Fig.~\ref{fig1}, $\hat{\Delta}_j$ corresponds to a pair of fermions between NN sites, i.e., an inter-CuO-chain pair. 
To avoid boundary effects, we set $j_0=L/4$ as the reference site.


\section{Results} \label{sec:results}

First, we show the pair correlation function $P(r)$ for various $J_1$ values. 
As shown in Fig.~\ref{fig2}, the pair correlations show power-law decays. 
$P(r)$ at $J_1/J_2=0.4$ exhibits a large magnitude and a slow decay.
However, the pair correlation decreases as $J_1$ decreases.  
All $P(r)$ values at $J_1/J_2=0.2$ and $0.4$ are positive, whereas the correlations at $J_1/J_2= -0.2$ and $0.0$ partially have negative values.  
These $J_1$-dependent properties suggest that the antiferromagnetic (ferromagnetic) $J_1$ regime is favorable (unfavorable) for spin-singlet SC. 

When the ground state is in the LE liquid phase, the charge density potentially shows a CDW-like signature~\cite{white2002,dolfi2015,lu2023,shen2023}.  
In Fig.~\ref{fig3}, we plot the local charge density $n(j)$ for various $J_1$ values. 
$n(j)$ at $J_1/J_2=0.4$ exhibits a clear density oscillation even around the center of the chain, where the wavenumber of the oscillation is consistent with $q = \pi n$.
The amplitude of the oscillation decreases as $J_1$ decreases.  
Similarly to the pair correlations, the CDW is unfavorable when $J_1$ is ferromagnetic.  

\begin{figure}[t]
\begin{center}
\includegraphics[width=0.935\columnwidth]{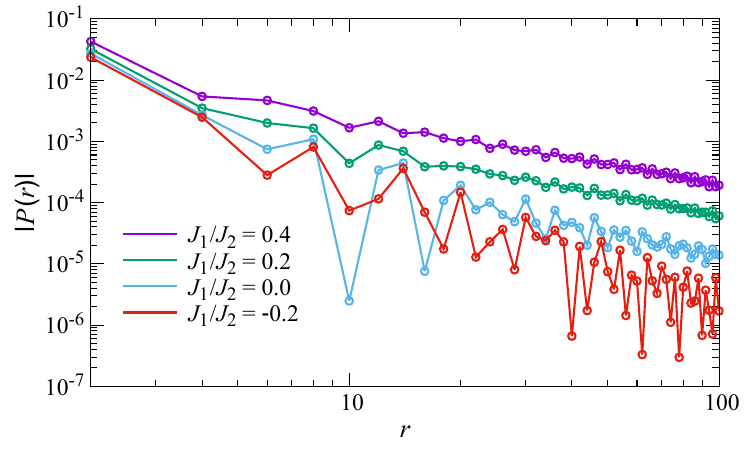}
\caption{Pair correlation function $P(r)$ for various $J_1$ values at $n=0.6$ filling, where the system size is $L=200$ and the reference site is $j_0 = L/4$.}
\label{fig2}
\end{center}
\end{figure}

\begin{figure}[t]
\begin{center}
\includegraphics[width=0.935\columnwidth]{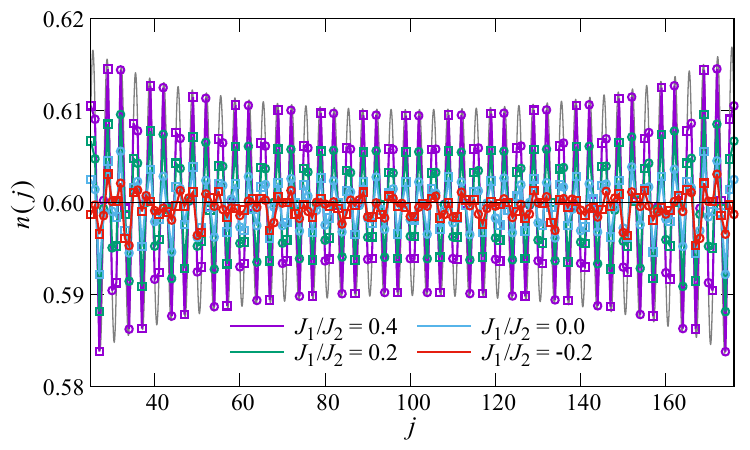}
\caption{Local charge density $n(j)$ for various $J_1$ values. 
Squares and circles indicate the densities at the odd (chain 1) and even (chain 2) sites, respectively.
The thin gray line is the fitting curve of Eq.~(\ref{eq:fig_charge}) for $J_1/J_2=0.4$. 
}
\label{fig3}
\end{center}
\end{figure}

The tendencies of the pair correlation $P(r)$ and the local charge density $n(j)$ suggest that the ground states at $J_1>0$ are in the LE liquid phase and its signatures gradually disappear as $J_1$ approaches zero. 
To collect the hallmarks of the LE liquid phase at $J_1 > 0$, we present other supporting quantities in Fig.~\ref{fig4}. 
In Fig.~\ref{fig4}(a), we plot the spin correlation function $S(r)$. 
At $J_1/J_2=0.4$, $S(r)$ exhibits an exponential decay, implying the presence of a spin gap, consistent with a property of the LE liquid (i.e., C1S0) phase.
To confirm the spin gap opening, in the inset of Fig.~\ref{fig4}(a), we show the spin gap $\Delta_{\rm s}(L) = E_0(L,S^z_{\rm tot}=1) - E_0(L,S^z_{\rm tot}=0)$, where $E_0(L, S^z_{\rm tot})$ is the lowest energy of the fixed system size $L$ with $S^z_{\rm tot}$. 
Extrapolating the computed data of $\Delta_{\rm s}(L)$ to the thermodynamic limit $L\rightarrow \infty$, we indeed find that the spin gap at $J_1/J_2 = 0.4$ is open.

\begin{figure}[t]
\begin{center}
\includegraphics[width=0.935\columnwidth]{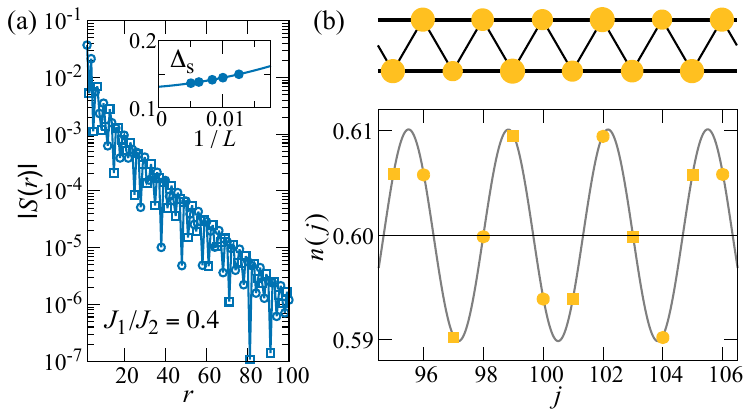}
\\
\includegraphics[width=0.935\columnwidth]{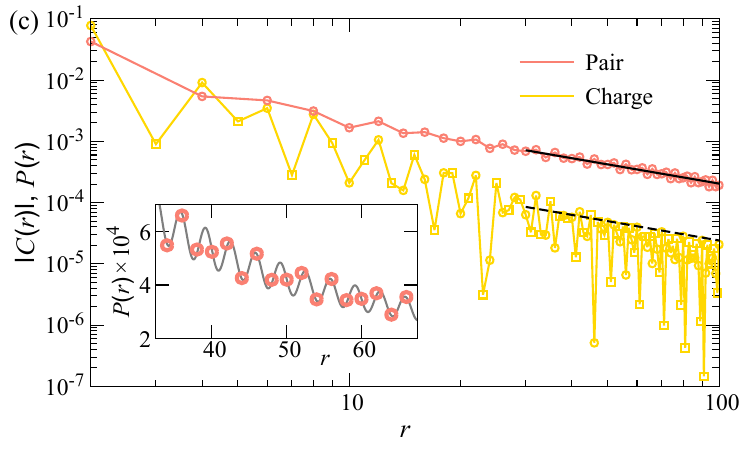}
\caption{(a) Spin correlation function $S(r)$ at $J_1/J_2 = 0.4$. 
Squares and circles indicate the correlations at odd and even distances, respectively, away from the reference site $j_{0} = L/4$. 
Inset: system size dependence of the spin gap $\Delta_{\rm s}(L)$ measured in units of $J_2$. The line denotes the second-order polynomial 
fit of $\Delta_{\rm s}(L)$. 
(b) Local charge density $n(j)$ around the center of the zigzag chain.  
The squares and circles in the lower panel of (b) indicate the densities at the odd (chain 1) and even (chain 2) sites, respectively.  
The upper figure of (b) schematically represents the charge density distribution on the zigzag chain. 
(c) Pair and charge correlation functions. 
The solid and dashed lines indicate the power-law decays of the pair and charge correlation functions, respectively. 
The inset shows the pair correlation function $P(r)$ and its fitting line. 
}
\label{fig4}
\end{center}
\end{figure}

The decaying behavior of the CDW correlation in the LE liquid phase is often evaluated using the local charge density $n(j)$~\cite{white2002,dolfi2015,lu2023,shen2023}. 
In Fig.~\ref{fig4}(b), we plot the calculated $n(j)$ and show a schematic figure of the charge density distribution around the center of the zigzag chain.  
We find that the charge density $n(j)$ at $J_1/J_2 = 0.4$ can be fitted by the function 
\begin{align}
n(j) = n + \delta n  \frac{\cos( \pi n j+\phi)}{\left[ \sin(\pi j /L) \right]^{K_c/2}}, 
\label{eq:fig_charge}
\end{align}
where $\delta n$ is the amplitude of the oscillating term and $\phi$ is the phase shift. 
$K_{c}$ may correspond to the decay exponent of the charge density correlation. 
Note that the fitting function of Eq.~(\ref{eq:fig_charge}) is modified from the function used in the two-leg ladder~\cite{white2002,dolfi2015} because we label the site index as in the 1D chain (see Fig.~\ref{fig1}). 
As shown in Fig.~\ref{fig3} (thin line) and Fig.~\ref{fig4}(b) (lower panel), the calculated $n(j)$ shows good agreement with the fitting curve of Eq.~(\ref{eq:fig_charge}). 
When the data at $50 < j \le 150$ are used in the fitting, we obtain $K_{\rm c} \simeq 1.05$.
In Fig.~\ref{fig4}(c), we compare the charge-density correlation function $C(r)$ with the decay line expected from the exponent $K_{\rm c}$. 
The decay of $C(r)$ in the long-range part (at $r\gg 1$) shows good agreement with the $r^{-K_{\rm c}}$ decay (black dashed line) when using $K_{\rm c}$ extracted from $n(j)$. 
Hence, $K_{\rm c}$ in Eq.~(\ref{eq:fig_charge}) gives a reasonable decay exponent for the correlation function $C(r)$. 

The decay exponent of the pair correlation function $P(r)$ is evaluated by the function 
\begin{align}
P(r) = \frac{A_0}{r^{\kappa_0}} + \frac{A_1 \cos(\pi n  r + \phi_1)}{r^{\kappa_1}}. 
\label{eq:fig_pairCF}
\end{align}
$A_0$ is dominant in $P(r)$, whereas a small $A_1$ gives a weak oscillation of $P(r)$. 
Indeed, as shown in the inset of Fig.~\ref{fig4}(c), the weak oscillation of the calculated $P(r)$ at $J_1/J_2=0.4$ is well fitted by the second term of Eq.~(\ref{eq:fig_pairCF}), where $\phi_1=\pi/2$ gives a reasonable fitting curve. 
Since the dominant contribution is attributed to the first term in Eq.~(\ref{eq:fig_pairCF}), $\kappa_0$ may correspond to the decay exponent of the pair correlation $K_{\rm sc}$~\cite{fabrizio1996,kuroki1997,dolfi2015,lu2023,shen2023}. 
When the data at $30 \le r \le 100$ are used in the fitting, we obtain $K_{\rm sc} = \kappa_0 \simeq 1.04$. 
Compared with the calculated $P(r)$ shown in Fig.~\ref{fig4}(c), the $A_0 r^{-K_{\rm sc}}$ term (black solid line) reproduces the decay of $P(r)$.  
As shown in Appendix~\ref{appendix_B}, the size dependence of $P(r)$ at $J_1/J_2=0.4$ is small, and thus the pair correlations may show the same decaying behavior even in larger systems. 
In the LE liquid phase, $K_{\rm c}$ and $K_{\rm sc}$ satisfy $K_{\rm c} K_{\rm sc}=1$~\cite{dolfi2015,lu2023,shen2023}. 
This relation is almost satisfied by $K_{\rm c}$ and $K_{\rm sc}$ extracted from our finite-size DMRG calculation at $J_1/J_2 = 0.4$. 
Hence, we conclude that the ground state of the zigzag chain at $J_1/J_2 = 0.4$ is in the LE liquid phase. 

\begin{figure}[t]
\begin{center}
\includegraphics[width=0.935\columnwidth]{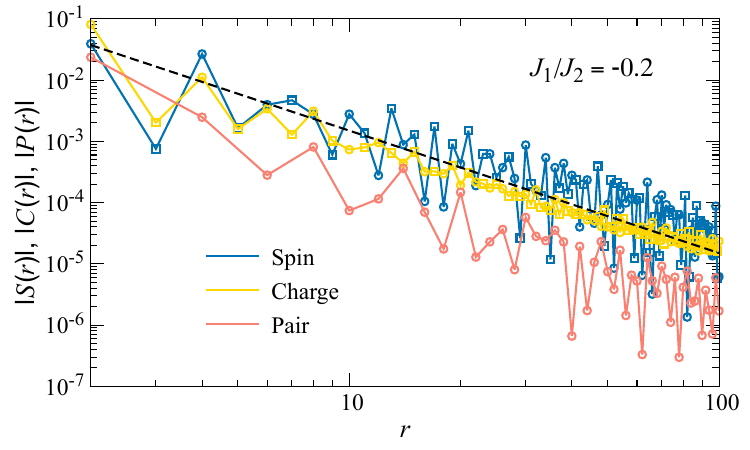}
\caption{Spin, charge, and pair correlation functions for $J_1/J_2=-0.2$.
The dashed (black) line indicates a $r^{-2}$ decay as a guide to the eye. 
}
\label{fig5}
\end{center}
\end{figure}

On the other hand, the decaying behaviors of the correlation functions at $J_1/J_2 = -0.2$ are different from those at $J_1/J_2 = 0.4$.   
In Fig.~\ref{fig5}, we plot the spin, charge, and pair correlation functions at $J_1/J_2 = -0.2$.  
In addition to the charge and pair correlation functions, the spin correlation function $S(r)$ also shows a power-law decay. 
As a guide to the eye, we plot a $r^{-2}$ decay line (dashed line) in Fig.~\ref{fig5}. 
The decays of all functions are close to the $r^{-2}$ decay. 
This may be a signature of a gapless Tomonaga--Luttinger liquid (i.e., both charge and spin sectors are gapless)~\cite{lu2023}. 
Therefore, the ground state in the ferromagnetic $J_1$ regime is no longer LE liquid and is unfavorable for the substantial development of the spin-singlet pair correlation.


\section{Summary and Discussion} \label{sec:summary}

We have investigated the correlation functions in the $t$-$J$ model for the CuO double-chain structure. 
In particular, we have focused on the roles of the NN, i.e., the interchain, spin exchange interaction $J_1$. 
Employing the DMRG method, we have demonstrated that the ground state for the antiferromagnetic $J_1$ shows the signatures of the LE liquid: the power-law decays of the pair and charge correlations and the exponential decay of the spin correlation. 
On the other hand, in the ferromagnetic $J_1$ regime, the pair correlation and charge density oscillation are suppressed, and the spin, charge, and pair correlation functions show similar power-law decays, implying that the ground state is no longer in the LE liquid phase. 

Our numerical demonstrations suggest that the antiferromagnetic NN interaction ($J_1>0$) favors SC in the CuO double-chain structure, e.g., in Pr$_2$Ba$_4$Cu$_7$O$_{15-\delta}$. 
To realize $J_1 > 0$, the antiferromagnetic contribution mainly originating from the electron exchange between the NN Cu sites must be larger than the ferromagnetic contribution mainly originating from Hund's coupling among the ligand $p$ orbitals.  
In the structure consisting of edge-sharing square-planar units, the direct $d$-$d$ hopping and detouring $d$-$p$-$p$-$d$ processes can contribute to the antiferromagnetic $J^{\rm (AFM)}_1$, whereas the contribution of the $d$-$p$-$d$ process to $J^{\rm (AFM)}_1$ is zero. 
However, if CuO$_4$ in the double-chain structure is distorted (as in Pr$_2$Ba$_4$Cu$_7$O$_{15-\delta}$), the $d$-$p$-$d$ process may contribute to $J^{\rm (AFM)}_1$. 
The quantitative estimation of $J_1$ including all spin-exchange processes in a realistic structure is an open issue for the future. 
As seen in our numerics in Fig.~\ref{fig4}(c), although the magnitude of $P(r)$ is larger than that of $C(r)$, the decay exponents of $P(r)$ and $C(r)$ are comparable ($K_{\rm sc} \sim K_{\rm c}$) at $r \gg 1$.  
These correlation properties and local density in Fig.~\ref{fig4}(b) imply that the SC pairing appears on the weak charge density modulation (i.e., SC coexists with CDW). 
This tendency can cause the suppression of the critical temperature in the CuO double-chain structure (e.g., $T_c \sim 20$~K in Pr$_2$Ba$_4$Cu$_7$O$_{15-\delta}$). 
A finding of an SC dominant regime without charge density modulation is an important open issue that should be addressed to obtain insights into high-temperature SC in the CuO double-chain structure.


\begin{acknowledgments}
We thank M. Ochi and T. Yagi for useful discussions. 
This work was supported by JSPS KAKENHI Grant Nos.~JP24K06939, JP24H00191 (T.K.), JP23K03286, JP24K01329 (K.S.), JP22K04907 (K.K.), JP20H01849 (T.K. and K.S.), and JP24K01333 (T.K. and K.K.) and by JST COI-NEXT Program Grant No. JPMJPF2221 (K.S.). 
The DMRG calculations were performed using the ITensor library~\cite{ITensor}. 
\end{acknowledgments}


\appendix 


\section{Total spin $S_{\rm tot}$ in the ground state}  \label{appendix_A}

\begin{figure}[b]
\begin{center}
\includegraphics[width=0.97\columnwidth]{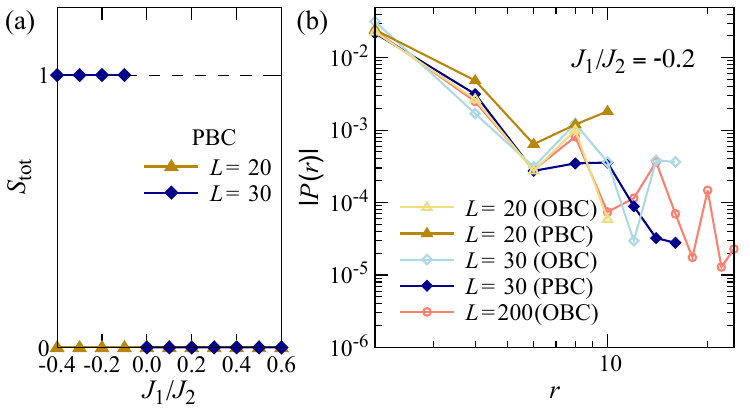}
\caption{(a) $S_{\rm tot}$ for $L=20$ and $30$ at $n=0.6$ under PBCs.
(b) Pair correlation function $P(r)$ at $J_1/J_2 = -0.2$ for various system size $L$ and boundary conditions. 
$j_0=7$ is used as the reference site when $L=30$, whereas $j_0 = L/4$ is used when $L=20$ and $L=200$. 
}
\label{figA1}
\end{center}
\end{figure}

The total spin in the ground state at $J_1 < 0$ (ferromagnetic) has the potential to be $S_{\rm tot} > 0$~\cite{doi1992}. 
Here, we examine the total spin $S_{\rm tot}$ obtained by calculating $\braket{\hat{\bm{S}}^2_{\rm tot}}=S_{\rm tot}(S_{\rm tot}+1)$. 
The ground states for $L=200$ with OBCs used in the main text have $S_{\rm tot} = 0$ even at $J_1/J_2 = -0.2$. 
However, $S_{\rm tot}$ at $J_1 < 0$ can be nonzero depending on the conditions in finite-size systems.  
In Fig.~\ref{figA1}(a), we plot $S_{\rm tot}$ as a function of $J_1$ for $L = 20$ and $L=30$ (at $n=0.6$) under periodic boundary conditions (PBCs).  
Note that the DMRG calculations under PBCs are more expensive than those under OBCs, and the largest truncation error in the calculations under PBCs is on the order of 10$^{-6}$ (when the bond dimension is up to $m= 10000$). 
As shown in Fig.~\ref{figA1}(a), $S_{\rm tot}=0$ for $L=20$, whereas $S_{\rm tot} > 0$ at $J_1/J_2 \le -0.1$ when $L=30$. 
Note that $S_{\rm tot}$ at $J_1/J_2=-0.2$ is also nonzero when $L=30$ under OBCs.  
Although the results are not presented, for $L=16$ and $L=32$ at $n=0.625$, $S_{\rm tot}>0$ under PBCs, whereas $S_{\rm tot}=0$ under OBCs at $J_1/J_2=-0.2$. 
In finite-size systems, $S_{\rm tot}$ at $J_1 < 0$ is sensitive to the system size, filling, and boundary conditions.   

Although $S_{\rm tot}$ has the potential to be nonzero at $J_1 < 0$, the small difference in $S_{\rm tot}$ in the ground state around $J_1 = 0$ may not improve the pair correlation. 
In Fig.~\ref{figA1}(b), we plot the pair correlation functions at $J_1/J_2=-0.2$ with different system sizes and boundary conditions, where $S_{\rm tot}=1$ when $L=30$ and $S_{\rm tot}=0$ otherwise. 
Although the accessible range is limited in small systems, the decaying behavior of $P(r)$ in the $S_{\rm tot}=1$ state ($L=30$) is not markedly different from that in the $S_{\rm tot}=0$ state. 
Hence, even in the $S_{\rm tot}=1$ state, the ferromagnetic $J_1$ regime is unfavorable for developing $P(r)$. 
This is consistent with our main conclusion.  
A large ferromagnetic $J_1$ ($<0$) may lead to the strongly spin-polarized ground state (with $S_{\rm tot} \gg 1$).
However, if $S_{\rm tot}$ is small around $J_1=0$, the ground-state properties are not markedly modified from those in the $S_{\rm tot}=0$ state.


\section{Size dependence of the pair correlation function in the LE liquid phase}  \label{appendix_B}

\begin{figure}[b]
\begin{center}
\includegraphics[width=\columnwidth]{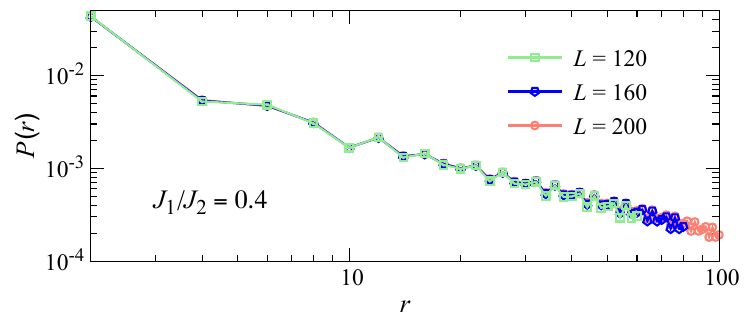}
\caption{Pair correlation function $P(r)$ at $J_1/J_2=0.4$ for $L=120$ (green squares), $L=160$ (blue pentagons), and $L=200$ (red circles) under OBCs.} 
\label{figB1}
\end{center}
\end{figure}

In Fig.~\ref{figB1}, we present the pair correlation function $P(r)$ at $J_1/J_2=0.4$ for various system sizes to examine a finite-size effect in the LE liquid phase.  
The deviation of the pair correlations between the smaller systems $(L<200)$ and the $L=200$ system used in the main text is barely changed, indicating the convergence of our numerical calculation. 
Therefore, the improvement of the decay exponent $K_{\rm sc}$ in larger $(L>200)$ systems is small. 
Although $K_{\rm sc}$ can vary depending on the choice of the system size, reference site, and fitting range of data~\cite{shen2023}, the results in the $L=200$ system used in the main text reasonably capture the characteristic of the LE liquid phase.


\bibliography{References}

\end{document}